\documentstyle[12pt,epsf]{article}
\input{epsf.tex}

\newcommand{\nc}{\newcommand}
\nc{\al}{\alpha}
\nc{\ga}{\gamma}   \nc{\Ga}{\Gamma}
\nc{\De}{\Delta}
\nc{\ald}{{\dot \al}}
\nc{\betad}{{\dot \beta}}
\nc{\gd}{{\dot \gamma}}
\nc{\sigmad}{{\dot \sigma}}
\nc{\mud}{{\dot \mu}}
\nc{\aldd}{{\ddot \al}}
\nc{\betadd}{{\ddot \beta}}
\nc{\gdd}{{\ddot \gamma}}
\nc{\sigmadd}{{\ddot \sigma}}
\nc{\mudd}{{\ddot \mu}}
\nc{\la}{\lambda}   \nc{\La}{\Lambda}
\nc{\var}{\varphi}  \nc{\hn}{h^\vee}
\nc{\pa}{\partial}  \nc{\hf}{\frac{1}{2}}         
\nc{\binomial}[2]{\left (\begin{array}{c} {#1}\\ {#2} \end{array}
\right )}
\nc{\ben}{\begin{equation}}
\nc{\een}{\end{equation}}
\nc{\bea}{\begin{eqnarray}}
\nc{\eea}{\end{eqnarray}}
\nc{\braket}[1]{\langle{#1}\rangle}

\nc{\C}{\mbox{\hspace{1.24mm}\rule{0.2mm}{2.5mm}\hspace{-2.7mm} C}}
\nc{\Nat}{\mbox{\hspace{.04mm}\rule{0.2mm}{2.8mm}\hspace{-1.5mm} N}}
\nc{\spa}{\hspace{1 cm},\hspace{1 cm}}
\nc{\vs}{\vspace}
\nc{\NP}[1]{Nucl.\ Phys.\ {\bf #1}}
\nc{\PL}[1]{Phys.\ Lett.\ {\bf #1}}
\nc{\CMP}[1]{Commun.\ Math.\ Phys.\ {\bf #1}}
\nc{\LMP}[1]{Lett.\ Math.\ Phys.\ {\bf #1}}
\nc{\PR}[1]{Phys.\ Rev.\ {\bf #1}}
\nc{\PRL}[1]{Phys.\ Rev.\ Lett.\ {\bf #1}}
\nc{\PTP}[1]{Prog.\ Theor.\ Phys.\ {\bf #1}}
\nc{\PTPS}[1]{Prog.\ Theor.\ Phys.\ Suppl.\ {\bf #1}}
\nc{\MPL}[1]{Mod.\ Phys.\ Lett.\ {\bf #1}}
\nc{\IJMP}[1]{Int.\ Jour.\ Mod.\ Phys.\ {\bf #1}}
\nc{\IM}[1]{Invent.\ Math.\ {\bf #1}}
\nc{\SJNP}[1]{Sov. J. Nucl. Phys.\ {\bf #1}}

      \def\Tr{{\rm Tr}}

\def\ni{\noindent}
\def\wt{\widetilde}      
      \def\ol{\overline}
\def\nn{\nonumber \\}

\def\bra#1{\,\left\langle\, #1 \,\right\vert}
\def\ket#1{\,\left\vert\, #1 \,\right\rangle}

\def\V{{\cal V}}
\def\U{{\cal U}}
      \def\H{{\cal H}}            
      \def\W{{\cal W}}            
      \def\LL{{\cal L}}           
                  
\def\N{{\cal N}}

\def\Z{{\bf Z}}

\def\C{{\bf C}}

\def\tq{{\tilde q}}

\begin{document}
\setcounter{page}{0}
\thispagestyle{empty}
\begin{flushright}
{OUTP}-01-13P\\
hep-th/0103064
\end{flushright}

\vspace{15mm}

\begin{center}
{\LARGE 
Boundary states
in boundary logarithmic CFT
}

\vspace{10mm}

{\Large
Yukitaka Ishimoto
\footnote{y.ishimoto1@physics.ox.ac.uk}
\\[10pt]
{\small 
Theoretical Physics, Department of Physics, \\
University of Oxford, \\
1 Keble Road, Oxford OX1 3NP, U.K.}
}
\end{center}

\vspace{10mm}

\begin{center}{\bf Abstract}\end{center}
{
There exist logarithmic CFTs(LCFTs)
 such as the $c_{p,1}$ models. 
It is also well known that it generally contains
Jordan cell structure.
In this paper, 
we obtain the boundary Ishibashi state for a $rank$-$2$ Jordan cell structure
and, with these states in $c=-2$ rational
LCFT, we derive boundary states in the closed string picture, which
correspond to boundary conditions in the open string picture. We also
discuss the Verlinde formula for LCFT and possible applications to string
theory. 
}

\vspace{35mm}

\ni
{PACS-codes: 11.25.Hf,
11.25.Sq,
11.10.Kk
}
\\
{
Keywords: logarithmic conformal field theory, boundary CFT, boundary states
}

\newpage
\baselineskip=17pt
\section{Introduction}

It is clear that 2d conformal field theory(CFT)\cite{BPZ} 
is an essential mathematical background to explore
string theories which are thought to be candidates of the long-awaited
ultimate theory of everything. 
Also, CFTs provide underlying theories or theoretical interpretations of 2-dimensional statistical physics.

It was first revealed by Rozansky and Saleur that some 4-point
functions of CFT have unavoidable logarithmic singularities\cite{roz},
and later, Gurarie showed that, with such logarithms, 
logarithmic fields appear in the theory, which was named, LCFT\cite{gura}.
The main feature of LCFT is that there is a pair, or maybe more, of
primary operators which are not independent 
and which form a reducible but indecomposable
representation of the $L_0$ operator,
$rank$-2 -- or even higher rank -- Jordan cell structure,
where one primary is logarithmic and the other is a state of zero 
norm\cite{CKT}.
In fact, 
some minimal models of CFT can have such fields in principle
although, in most cases, they are non-unitary or central charges of them
are, somehow, irregular. 
Nevertheless,
it is worth seeing the extent to which we can investigate them
for new physics based on them,
since we can ignore their non-unitary nature by having them as subsystems. 
Thus far, many studies have been devoted to this subject and have found
the same sort of logarithmic behaviour in various models.
For example, 
the gravitationally dressed CFT and 
WZNW models at different levels or on different 
groups[2, 5-7],
$c_{p,1}$ and non-minimal $c_{p,q}$ models, as mentioned above,
$c_{2,1}=-2$ 
model[3, 8-15],
$c=0$ models,
describing critical polymers and percolation\cite{salu,kau3,card7,perco},
quantum Hall effect,
quenched disorder and localisation in planar systems\cite{qhe},
2D-magneto-hydrodynamic and ordinary turbulence\cite{rahimi1}.
In string theory, D-brane recoil, target-space symmetries and $AdS/CFT$
correspondence have been studied and discussed with respect to LCFTs in
the literature\cite{kog1,ads3}.

On boundary CFT, which is CFT with one or more boundaries and
boundary conditions,
it was shown by Cardy that, with the boundary conditions, a lot of the
tools developed in ordinary CFT can be used and hence 
n-point functions become manageable \cite{card1}.
These types of theories are
essential
in both particle physics and condensed matter physics,
when some direction is required to be finite or to have one or two ends.
For instance, in string theory, theories of open strings are defined on an
infinite strip with two boundaries.
A periodicity along its boundaries induces dual pictures on it and modular
invariance leads to a one-to-one correspondence between the boundary
conditions of the open string picture and the boundary states of the closed 
string picture, which is, on the other hand, quantised on an annulus.
It was found in \cite{ishibashi1} that these boundary states are spanned by
boundary {\it Ishibashi} states, by which the Verlinde formula is proven to
hold for unitary minimal models of boundary CFT\cite{card1,verl1}.

In spite of much progress in both areas, little has been
mentioned on LCFT with boundaries and the effects of the presence of
boundary, because there is a problem of reducible but indecomposable
representations which cannot be applied to boundary CFT in a
straightforward way.
The first systematic attempt to formulate boundary LCFT was made by
Kogan and Wheater in \cite{kog3}, where several important problems were 
discussed, including the structure of boundary states in LCFT, using the
$c=-2$ theory as an example, and the Verlinde formula.
The arguments on boundary states are based on the conjectured forms of
the Ishibashi states and therefore the conjecture remains to be proven.
Otherwise, it should be confirmed that we can derive explicit forms of
Ishibashi states without relying on it, and whether they reproduce the
same result.

In this paper, 
we briefly review boundary CFT, LCFT and the definition of Jordan cell
structure.
Thereafter, 
we prove the existence of the boundary Ishibashi state for
the $rank$-$2$ Jordan cell structure and show explicit forms of them.
We also propose a conjecture of Ishibashi states for all LCFTs which contain
$rank$-$2$ Jordan cell structure. 
After introducing the $c=-2$ LCFT,
we show how these states prescribe the boundary states in the closed
string picture, which correspond to boundary conditions in the open
string picture. 
In consequence, we show some typical results, which potentially
include the one given in \cite{kog3}, and take the
different original result as a conclusion.
Finally, we will also discuss the Verlinde formula for LCFT and 
possible applications to string theory. 

\section{Preliminaries}
\subsection{Boundary CFT}

To begin with,
consider an infinite strip of width $L$ on which theories of
open strings can lie. 
By a conformal map,
$
w = \frac{L}{\pi} \ln z 
$, 
a theory on a $z$ upper half-plane
is mapped onto a $w$ infinite
strip, where time $t$ goes
along two parallel edges.
A pair of conformally invariant boundary conditions is put onto these two
edges respectively, labeled by $\al, \beta$, and the Hamiltonian of this
system is given by $(\pi/L)H_{\al \beta}$ with a generator of
$t$-translations, $H_{\al \beta}$. 
The eigenstates of $H_{\al \beta}$
fall into irreducible representations of the chiral algebra and the partition
function becomes a linear combination of the functions of these representations. 
By imposing a periodicity $T$ along $t$, 
the partition function of the open string picture reads
\bea
 \label{def:Zopen}
 Z_{\al \beta}^{open} (q)
= \Tr\, q^{H_{\al \beta}}
= \sum_{i} n_{\al \beta}^{\;\; i} \, \chi_i (q) , 
\eea
where $q\equiv e^{2\pi i \tau}, \tau\equiv i T/2L$ and $n_{\al \beta}^{\;\; i}$
is the number of times which a representation $i$ occurs in the presence
of boundary conditions $(\al \beta)$. $\chi_i (q)$ denotes a character
function of the representation $i$.

At this point, a dual picture appears.
The periodicity wraps the strip to a cylinder
and the dual description of the theory is given by the change of $t$
direction to one across the strip. 
The boundary conditions turn to the
boundary states on the initial and final ends of the cylinder and the
partition function
of the dual picture is constructed from the theory of
closed strings. This cylindrical geometry, an annulus on $\zeta$-plane, 
is obtained from the strip by a map,
$w = i \frac{T}{2 \pi} \ln\zeta$, 
and the conformal invariance of the boundary conditions amounts to
the following conditions of the
boundary states, $\{\ket{B}\}$:
\bea
 \label{def:bc0}
   ( W_n - (-1)^s \ol{W}_{-n} ) \ket{B} &=& 0 ,
\eea
where $W_n$($\,{\ol{W}_n}\,$) denotes a $n$-th mode of
the (anti-)holomorphic sector of the chiral algebra, and $s$ is a dimension
of the operator. 
Among solutions of eq.(\ref{def:bc0}), {\it Ishibashi} states are known to
form a basis of boundary-state space and 
express the partition function of the closed string picture in a more
convenient way as below.
\bea
 \label{def:Zclose}
   Z_{\al \beta}^{closed} (\wt q) 
= \bra{\wt\al} \wt q^{\frac12 \left( L_0 + \ol{L}_0 - \frac{c}{12}
\right)} \ket{\wt\beta} 
= \sum_{i} \braket{\wt\al|i}\braket{i|\wt\beta} \chi_{i}(\wt q),
\eea
where $\wt q\equiv e^{2\pi i \wt\tau}, \wt\tau\equiv - 1/\tau$.
Note that $\{\ket{i}\}$ denote Ishibashi states and the
diagonality of them, that is, of the representations is used. 
The same central charge is assigned to both chiral sectors, i.e. $\ol{c}=c$.

As a consequence, the equivalence of both quantisation schemes ends up with 
\bea
 \label{def:equiv-dual}
   Z_{\al \beta}^{open} (q) =  Z_{\al \beta}^{closed} (\wt q) ,
\eea
where both sides of the equation have the same set of characters but of
different variables. Since $\wt\tau=-1/\tau$, modular properties of the characters 
lead to the relations between boundary states and 
$n_{\al \beta}^{\;\; i}$, by which  
the forms of the boundary states, in terms of Ishibashi states, 
and the values of $n_{\al \beta}^{\;\; i}$ are equated.
Actually, modular properties of characters completely determine the
above quantities in unitary minimal models and 
lead to the Verlinde formula of boundary CFT, 
provided that $n_{\al \beta}^{\;\; i}=\delta_\beta^i$ for some $\al$. 
The condition is satisfied when the theory has a `vacuum' and 
$n_{\al \beta}^{\;\; i}$ is identical to the fusion rule
coefficients $\N_{\al \beta}^{\;\; i}$ of the theory. 
Since the identification is precisely what the formula means, 
this should be taken as the self-consistency condition of the formula,
which is not sufficient.

In addition,
the diagonality of Ishibashi states is essential in this construction,
and this seems to be absent in LCFT, since LCFT possesses reducible but
indecomposable representations which are obviously {\it not} diagonal. 
Nonetheless, there might be a possibility that Ishibashi states of LCFT 
allow the similar construction and hence the Verlinde formula.
This is worth being carefully examined.

Before we turn to the Ishibashi states, it is better to see what LCFT
is like and how initial states of Jordan cell structure can be defined.
Being based on them, 
detailed proof and examinations of Ishibashi states
will be given in the next section.

\subsection{Jordan cell \& LCFT}

In unitary minimal models, the theory is characterised by a central
charge $c_{p,q}= 1 - 6 \frac{(p-q)^2}{p q}$ and conformal dimensions of
the fields $\phi_{r,s}(z)$, $h_{r,s}=\{(rp-sq)^2-(p-q)^2\}/4pq$, where
$p\geq 2, q=p+1$ are 
integers and the set of integers $(r,s)$ is restricted to a
rectangular region, $1\leq r<q, 1\leq s<p$. 

If we remove the constraint on $(p,q)$,
the $c_{p,1}$ models appear as non-unitary theories, 
where the above rectangular
regions vanish and so do the restrictions on $(r,s)$.
Instead, due
to the relations $h_{r,s}=h_{-r,-s}=h_{r+1,s+p}$, the region for $(r,s)$
is stretched to a semi-infinite rectangular region $1\leq r ,\, 1\leq s\leq p$.
Remarkably, fusion rules of them result in 
distinguishable states of the same conformal dimension and hence
degenerate theories. They might be simply degenerate and diagonalisable,
but in fact, in $c_{2,1}=-2$ model, a logarithmic field $D(z)$ emerges
in the fusion rule of the primary $\mu(z)\equiv\phi_{2,1}(z)$ and gives a
logarithmic singularity in its 4-point function. 
This, together with a normal primary field $C(z)$ of the same
dimension, forms a reducible but indecomposable representation.  
The general form of the pair of such fields can be written down as
\begin{eqnarray}
 \label{def:CD}
  T(z) C(w) &\sim& \frac{h \,C(w)}{(z-w)^2} + \frac{\partial_w C(w)}{z-w} , \nn
  T(z) D(w) &\sim& \frac{h \,D(w) + C(w)}{(z-w)^2} + \frac{\partial_w D(w)}{z-w} ,
\end{eqnarray}
where $h$ is a conformal dimension of both fields and 
their correlation functions are given by \cite{gura,CKT}
\begin{eqnarray}
 \label{def:rank2JS}
  \braket{C(z) C(w)} &\sim& 0 , \quad
  \braket{C(z) D(w)} \,\sim\, \frac{\al}{(z-w)^{2 h }} , \nn
  \braket{D(z) D(w)} &\sim& \frac{1}{(z-w)^{2 h }}\left(-2 \al \ln (z-w) + \al^{\prime} \right) .
\end{eqnarray}
Accordingly, a pair of initial states $\ket{C}$ and $\ket{D}$ forms a $rank$-2
Jordan cell,
\begin{eqnarray}
 \label{def:rank2}
  L_0 \ket{C} = h \ket{C} ,\quad L_0 \ket{D} = h \ket{D} + \ket{C} . 
\end{eqnarray}
Verma modules of them are obtained from the above states by acting
with the chiral algebra successively on them.

\subsection{Jordan cell structure}

On the way to consistent boundary LCFT, we fix the notations of Jordan
cell structure,
most of which has been introduced by Rohsiepe.\cite{roh1}

Let $\U$ be the universal enveloping algebra of Virasoro algebra $\LL$.
Setting $\LL^{\pm}\equiv < L_{n \!{\tiny \begin{array}{l} >\\[-4pt]< \end{array}} \!0 } >$,
$\LL^0\equiv < L_0 , C >$, we can introduce $\U^{\pm},\U^0 \subset \U$, 
the enveloping algebras of them. Note that this can be naturally extended to
any chiral algebras that are graded in the same way.

Jordan lowest weight module(JLWM) is defined by Rohsiepe as a ${\cal
L}$-module, $\V$, satisfying 
\bea
 \label{def:JLWM}
   &(0)& C v^{(i)} = c v^{(i)} , \nn
   &(1)& L_0 v^{(i)} = h v^{(i)} + v^{(i-1)} ,
    \quad   L_0 v^{(0)} = h v^{(0)} ,\quad ( h,c \in{\cal C} ) \nn
   &(2)& v^{(i)} \in \V_0 \qquad ( 0 \leq i \leq k-1 ),  \nn
   &(3)& \V = \U . v^{(k-1)} , 
\eea
where $\V_0 \equiv \{v\in\V|{}^\forall v^{\prime} ; {\cal U}^+ v = 0,
v\not=\U^- v^{\prime} \}$, $h$ is lowest weight, $\{ v^{(i)} \}$ are
linearly independent lowest weight vectors(JLWV). The integer $k$
is called $rank$ of JLWM. For $rank$-2 case, $v^{(0)}$ is called upper
 JLWV and $v^{(1)}$ is lower JLWV.

One may define a representation of $\LL$ on its dual $\V^*$ by setting 
\bea
 \label{def:*}
\left( L_{i_1}^{n_1} \cdots L_{i_p}^{n_p}
 \right)^\dag
&=&
  L_{-i_p}^{n_p} \cdots L_{-i_1}^{n_1}, \nn
   \left( \LL \phi \right) (w) &=& \phi (\LL^\dag w) {\rm ~~for~\phi\in
   \V^* ,~} w \in \V .
\eea
$\V^*$ appears as a JLWM with lowest weight vectors, $v^{(i) *}$, which
satisfy eq.(\ref{def:JLWM}). Therefore, the dual JLWM, $\V^\dag \subset
\V^*$, is naturally induced by a map,
$
 \label{def:dual}
  \V=\{u.v\} \rightarrow \V^\dag=\{u.v^*\},
$
where $u\in\U$, $v,v^*$ are lowest weight vectors on each side, respectively.

The Shapovalov bilinear form, $\braket{ ~|~ }$, can be defined as\footnote{You may change this orthogonality condition for some cases.}
\bea
 \label{def:shapo}
   {}^\forall v^{(i)} \in \V, ~{}^\exists v^{(j) \dag}\in\V^\dag ~;  \quad
   \braket{v^{(j)}|v^{(i)}} \,=\, \delta_{ij}
\eea
which, of course, satisfies $\braket{v_i |L_n^{\dag}|v_j} = \braket{v_i
| L_{-n} |v_j }$ and, in fact, this condition prescribes the relation
between $*$ and $\dag$. Namely, the $*$ transformation is an isomorphism
which acts on $\V$ as
\bea
 \label{def:*JLW}
   * : v^{(i)} &\rightarrow& v^{(i) *} = v^{(k-1-i) \dag} ,
\eea
therefore, $ \ket{v^{(i)}} \rightarrow \bra{v^{(k-1-i)}} ,
       \bra{v^{(i)}} \rightarrow \ket{v^{(k-1-i)}}$.\footnote{
Note that, even if you have
$\braket{v^{(0)}| v^{(1)}} \not= 0$
for $rank$-2 case, we have the same result. 
}

On $\V$, Virasoro generators can be divided into two parts, namely,
$L_n = L_n^d + L_n^n$, such that,
\bea
 \label{def:d+n}
    L_n^d &:& \ket{v^{(i)}, N} \rightarrow \ket{v^{(i)}, N-n} , \nn
    L_n^n &:& \ket{v^{(i)}, N} \rightarrow \ket{v^{(i-1)}, N-n} \quad
    for ~i\not= 0 , \nn
   &&  \ket{v^{0}, N} \rightarrow  0 ,
\eea
where, for simplicity, $\ket{v^{(i)}, N}$ denotes orthogonal basis of
$\ket{v^{(i)}}$-descendants at level $N$. We will use this convention
from now on.

\section{Boundary Ishibashi states}

It is natural to assume that a solution $\ket{B}$ of eq.(\ref{def:bc0})
takes the form,
\ben
 \label{def:B}
\ket{B} \equiv \sum_{\{N\}} \ket{ \al \,,\,N }\otimes\ol{\ket{ \beta \,,\,N}},
\een
because initial(final) states are in the tensor product of Hilbert spaces
of both chiral sectors, $\H \otimes \ol\H$.
Similarly, eq.(\ref{def:bc0}) is equivalent to
\bea
 \label{calc}
   \bra{j,N_1}\otimes\ol{\bra{k,N_2}} ( W_n - (-1)^s \ol{W}_{-n} )
   \ket{B} &=& 0 , 
\eea
where 
$j, k, N_1, N_2$ are arbitrary.

Extracting Virasoro parts of the above conditions, let us solve the equation
\bea
 \label{eq:L}
   \bra{j,N_1}\otimes\ol{\bra{k,N_2}} ( L_n - \ol{L}_{-n} ) \ket{B}
 &=& 0 , 
\eea
where the left hand side can be decomposed and simplified into two parts, 
according to the decomposition of Virasoro algebra
in the previous section.
\bea
 \label{calc:L}
   lhs &=& \sum_{N} \Bigg\{ \braket{j,N_1|(L_n^d + L_n^n)|\al,N}
 \braket{ \ol{k,N_2}|\ol{\beta,N} }
\nn
&&\qquad\qquad - \,\,\braket{j,N_1|\al,N} 
 \braket{ \ol{k,N_2}|(\ol L_{-n}^d + \ol L_{-n}^n)|\ol{\beta,N}} \Bigg\} 
\nn
   &=& ({\rm diagonal~part}) + ({\rm non{-}diagonal~part}),
\nn
({\rm diagonal~part})&\equiv& \delta_{N_1,N_2-n} \delta_{k,\beta} \delta_{j,\al}
 \Big\{ \bra{\al,N_1} L_n^d \ket{\al,N_1 + n} - \bra{\beta^*,N_1}
  L_{n}^d \ket{\beta^* , N_1+n} \Big\} ,
\nn
({\rm non{-}diag~part}) 
&\equiv& \delta_{N_1,N_2-n} \left( \delta_{j,\al} \delta_{\beta^* < k^*}
 + \delta_{j<\al} \delta_{\beta^*,k^*}
 + \delta_{j<\al} \delta_{\beta^* < k^*} \right)
\nn 
&&  \times  \bra{j,N_1} \Big( L_n \ket{\al,N_2} \bra{\beta^*,N_2} -
     \ket{\al,N_1} \bra{\beta^*,N_1}L_n \Big) \ket{k^*,N_2} . 
\eea
where we introduce
$\delta_{f>i}$ such that, if $
f=v^{(a)}, i=v^{(b)}$ and $a>b$, then $\delta_{f>i}=1$, others vanish.
Here we assume that the above bilinear form is a Shapovalov form but
not necessarily simple under $*$, and $\braket{j,N_1|L^n_n|\al,N_2}\not=0$
if and only if $h_j = h_{\al}$.
The diagonal part vanishes if $\ket{\al,N}$ and $\ket{\beta^*,N}$ have
the same conformal structure, i.e. the same conformal dimensions and null
 vectors at the same level, etc. For the case of Sugawara construction,
 this is compensated by setting two states to be in the same multiplet 
as
$\ket{\al,0}=\ket{0;l,m}$ and 
$\ket{\beta^*,0}=\ket{0;l,m^{\prime}}$\cite{ishibashi1}. However, for
 $c_{p,1}$ models, there is a
 possibility for $\ket{\alpha}$ and $\ket{\beta}$ to be not in the same 
representation but in the same Jordan cell.

Given that we have just two primary states in a
$rank$-$2$ Jordan cell, namely, the upper JLWV $\ket{C}=\ket{v^{(0)}}$ and
the lower JLWV $\ket{D}=\ket{v^{(1)}}$, while
generators of Virasoro algebra merely generate their descendants,
i.e. both submodules, $\V_C$ and $\V_D$, of JLWM do not contain any other submodule. 
By forcing $\al, \beta$ to be in
this cell, vanishing diagonal part is assured and corresponding
conditions for boundary Ishibashi states
reduce to 
\bea
 \label{eq:nonW}
&&   ({\rm non{-}diag~part}) = 0  \qquad{\rm
   for~arbitrary~}j,k,n,N_1,N_2 ,  
\nn
   lhs &=& \delta_{N_1,N_2-n} \Bigg[
 \delta_{j,\al} \delta_{k,C} \delta_{\beta,D}
    \Big( \bra{\al,N_1} L_n \ket{\al,N_2} \bra{C,N_2} -
    \bra{C,N_1}L_n \Big) \ket{D,N_2} 
\nn
&&{}+ \delta_{\beta,k} \delta_{j,C} \delta_{\al,D}
    \bra{C,N_1} \Big( L_n \ket{D,N_2}  -
     \ket{D,N_1} \bra{\beta^*,N_1} L_n \ket{\beta^*,N_2} \Big) 
\nn
&&{}+ \delta_{j,k,C} \delta_{\al,\beta,D}
    \braket{C,N_1|L_n|D,N_2} \Big( \braket{C,N_2|D,N_2} -
     \braket{C,N_1|D,N_1} \Big) \Bigg]
. 
\eea
Then finally we get the following conditions, 
\ben
 \label{eq:nonW-bc}
\delta_{\al,D} = \delta_{\beta,D} = 0.
\een
Hence, the only allowed Ishibashi state in the Jordan cell is, as expected,
\ben
 \label{def:bs-0}
   \ket{B} = \sum_{\{N\}} \ket{C, N} \otimes 
   \ol{\ket{C, N}} . 
\een
This new result is
valid for all $rank$-$2$ indecomposable representations of this type, 
as long as both $\V_C$ and $\V_D$ have the same conformal
structure and spectrum.
Unless the assumption is violated, we can extend the chiral algebra as
far as possible.
Unfortunately, this is not the case of the $c=-2$ rational LCFT, because 
the conformal tower of the logarithmic state $\ket{\omega}$ contains a
subrepresentation.\footnote{
$\ket{\omega}\equiv\ket{D,h=0}$, $\ket{\Omega}\equiv\ket{C,h=0}$. For
instance, $\ket{\phi} \equiv L_{-1} \ket{\omega}$ should be
interpreted as in a subrepresentation.
}
However, it is still a rigorous proof of boundary states in Jordan cell 
structures and
it could be extended to a generic case. 

Namely, we state a conjecture 
that, in general, for all $rank=2$ indecomposable representations, only one 
boundary Ishibashi state is allowed in each representation.

If all LCFTs contain Jordan cell structure, this result and 
conjecture become a powerful fundamental tool to tackle to boundary LCFT.
Also, it should be noted that this
conjecture includes $W$-algebraic cases, 
which are thought to give rational series of $c_{p,1}$ LCFTs. 

\section{$c=-2$ boundary LCFT}

At this stage, we can investigate boundary LCFTs which only
contain $rank$-2 Jordan cell structures and irreducible representations.
A useful point in these LCFTs is that the number of independent boundary
states coincides with 
the number of Jordan cells plus the number of irreducible
representations\footnote{
e.g. if there are one Jordan cell and one irreducible representation,
there are two independent boundary states because of two independent
Ishibashi states.
}.

In this section, we will show how we can obtain boundary states in a
particular LCFT, the
$c=-2$ theory, as a first example.
In order to do so, we briefly review the $c=-2$ theory in
\cite{gab3,flohr1,kog3}
from the point of view of modular properties. Then we examine the
boundary states of the theory.

\subsection{The $c=-2$ again}

The Jordan cell in eq.(\ref{def:rank2}) sits on the vacuum representation,
and two Verma modules of the cell are of the same sort. 
In other words, the characters of them take the same form and 
become indistinguishable.
It may be possible to interpret them as two coincident characters of 
different representations.
When the chiral algebra of the system is Virasoro algebra,
characters of the theory are given straightforwardly
as Virasoro characters. 
The rest of our construction
seems to be rather easy, but it is to be taken carefully. 
In fact, the chiral algebra should include $\W$-algebra.

In boundary CFTs on a wrapped strip,
the partition function as a sum of
characters has to be transformed into a sum of the same set of 
characters under modular transformations.
Therefore, at least, we need some sort of rationality
which plays a crucial role in unitary rational models.
Since above mentioned Virasoro
characters in `normal', i.e. without $\W$-symmetry, $c=-2$ model
are not 
modular-transformed into the same set but generate 
more characters, 
the rationality is missing.
However, there is another way to recover the rationality, that
is, with $\W$-symmetry. Precisely, the number of representations
is reduced to be finite and a set of linear combinations of them may possess
well-defined modular properties.\cite{gab3,flohr1}
This is why we are about to take it in our theory.
It should be noted that the loss of rationality always happens 
in normal $c_{p,1}$ models.

In $c=-2$, $\W(2,3,3,3)$ algebra plays this role, representations of which
are summarised as $h=\{-1/8,0,3/8,1\}$\cite{gab3}. 
Their $\W$-characters are given by
\begin{eqnarray}
 \label{def:allchara}
   \chi_{\V_0} ( q )
   &=& \frac{1}{2\,\eta(q)} \left( \Theta_{1,2}(q) + \pa\Theta_{1,2}(q) \right) , \nn
   \chi_{\V_1} ( q )
   &=& \frac{1}{2\,\eta(q)} \left( \Theta_{1,2}(q) - \pa\Theta_{1,2}(q) \right) , \nn
   \chi_{\V_{-1/8}} (q) 
   &=& \frac{1}{\eta(q)} \Theta_{0,2}(q) ,\quad
   \chi_{\V_{3/8}} (q) 
   \,=\, \frac{1}{\eta(q)} \Theta_{2,2}(q) ,
\end{eqnarray}
where $q=e^{2\pi i \tau}$, $\Theta_{l,k}(q)\equiv\sum_{n\in\Z} q^{(2kn + l)^2/4k}$ 
is a Riemann theta function.
 and
$\pa\Theta_{l,k}(q)\equiv\sum_{n\in\Z}
\left(2kn+l\right) q^{\left( 2kn+l \right)^2 / 4k}$. 
The first character is of the vacuum representation and of the Jordan cell.
Characters in eq.(\ref{def:allchara}) do not close under modular
transformations but generate the new function $\Delta\Theta_{1,2}/\eta\equiv
i\tau\pa\Theta_{1,2}/\eta$. In fact, linear combinations of
those five functions can form a modular invariant set.

One way is to introduce the notion of generalised highest weight representations(hwrep), $R_0$
and $R_1$, and define $\chi_{R_0} = \chi_{R_1} =
2\left(\chi_{\V_0}+\chi_{\V_1}\right)$ so that, 
with $\chi_{\V_{-1/8}},\chi_{\V_{3/8}}$, they form a modular invariant set.
This was first proposed in \cite{gab3}, 
based on the analysis of fusion rules,
and its $S$-matrices are given in \cite{gab3,roh1}.

Another way is to draw a general set of linear combinations and
determine the coefficients so that
the modular invariant partition function is given by them.
Separately, this was given in \cite{flohr1}, 
showing that there are three cases.
$S$-matrices of them are also listed.

The way which has been taken in \cite{kog3}
is to define the logarithmic pair by two non-logarithmic primaries in
a particular limit and infer their characters should be those
in \cite{flohr1}.
Selecting four characters, the $S$-transformation is expressed by
two different matrices, $S$ and $Q$, the latter of which is for
the logarithmic nature. They imposed invariance under $S^2$-transformation and
derived the result.

In what follows, we will start from the first way, and then, case(I)
in \cite{flohr1} and check the last one. Note that
the first two ways are in the scope of LCFT without boundaries and one of
necessary conditions is $S^4=1$, which should be $S^2=1$ in 
boundary cases.

\subsection{Boundary states and `fusion rule' coefficients}

\subsubsection{First approach: on Gaberdiel and Kausch's construction}

In the first approach of the $c=-2$ LCFT, there are two generalised hwrep,
$R_{0}$, $R_{1}$, 
and two normal lwrep, $\V_{-1/8}$, $\V_{3/8}$, 
where two different
logarithmic pairs reside in each generalised one. 
Therefore, we have four linearly independent boundary Ishibashi
states, two of which are constructed from upper JLWVs as in the previous
section. Fortunately, they are orthogonal to each other and span the
space of boundary states. 

The main aim of this section is to construct the set of boundary
states from them, which correspond to boundary conditions in
the open string picture. We then discuss the 
Verlinde formula for the LCFT.

Let us begin with the closed string picture, that is, LCFT on a
cylinder. Extracting upper JLWVs, $\Omega$ and $\phi$,\footnote{
The upper JLWV $\phi_{\al}$ in $R_1$ is a doublet
under the $\W$-algebra, but we suppress the suffix for simplicity.
}
out of $R_{0,1}$, any initial boundary state is expressed as 
$$ \ket{intial~state} = a \ket{\Omega} + b \ket{\phi}
 + c \ket{-\frac{1}{8}} + d \ket{\frac{3}{8}} ,$$
and final state is done similarly, where each bra(ket) in $r.h.s.$
denotes an Ishibashi state. 
As there are supposed to be four boundary conditions for
the open string picture, the corresponding boundary states may be labeled
by $\wt\al = \{\wt{R_0}, \wt{R_1}, \wt{-1/8},\wt{3/8}\}$ and are given by
\bea
 \label{def:boundary state}
   \ket{\wt \al} &\equiv&  \sqrt{2}\, \bar{a}_{\al} \ket{\Omega}
   + \sqrt{2}\, \bar{b}_{\al} \ket{\phi}
   + \bar{c}_{\al} \ket{-\frac{1}{8}} + \bar{d}_{\al} \ket{\frac{3}{8}},
\nn
   \bra{\wt \al} &\equiv&  \sqrt{2}\, {a}_{\al} \bra{\Omega}
   + \sqrt{2}\, {b}_{\al} \bra{\phi}
   + {c}_{\al} \bra{-\frac{1}{8}} + {d}_{\al} \bra{\frac{3}{8}} ,
\eea
with a factor $\sqrt{2}$ added for later convenience.

With the definition of Ishibashi states, eq.(\ref{def:B}), 
the partition function becomes simple in terms
of the characters:
\bea
 \label{eq:Zcyl}
   Z_{\al \beta}(\tq) 
&=& \braket{\wt{\al}|\tq^{\frac12 \left( L_0 + \wt{L}_0
     - c/12 \right)}|\wt{\beta}} 
\nn
&=& 2 \left( a_{\al} \bar{a}_{\beta} \chi_{\V_0}(\tq) + b_{\al}
\bar{b}_{\beta} \chi_{\V_1}(\tq) \right) + c_{\al} \bar{c}_{\beta}
\chi_{\V_{-1/8}}(\tq) + d_{\al} \bar{d}_{\beta}
\chi_{\V_{3/8}}(\tq) .
\nn
&=& a_{\al} \bar{a}_{\beta} \left\{ \ga \chi_{R_0}(\tq)
 + \left(1-\ga\right) \chi_{R_1}(\tq) \right\}
 + c_{\al} \bar{c}_{\beta} \chi_{\V_{-1/8}}(\tq)
 + d_{\al} \bar{d}_{\beta} \chi_{\V_{3/8}}(\tq)
\nn
&=& (M_{\al})_{\beta}^{\;\; j} \,\chi_j (\tq)  ,
\eea
where $\tq\equiv e^{2\pi i \wt\tau} = e^{-2\pi i /\tau}$, $j$ is
contracted and summed
over $R_0, R_1, \V_{-1/8}$, and $\V_{3/8}$. 
In the second line of eq.(\ref{eq:Zcyl}), the first two
coefficients are naturally {\it combined} into one with a condition 
$a_{\al}\bar{a}_{\beta} = b_{\al}\bar{b}_{\beta}$, due to the modular
invariance of the theory. In addition, due to $\chi_{R_{0}}=\chi_{R_{0}}$,
lack of difference between $\chi_{R_{0,1}}$, 
an extra factor $\ga$ is introduced to {\it
redistribute} the
combined term to two characters, which emerge in the open string picture.
Finally, the form of the partition function is simplified with a
matrix 
$M_{\al}=\left(a_{\al} \bar{a}_{\beta} \ga, 
         a_{\al} \bar{a}_{\beta} \left(1-\ga\right),
	 c_{\al} \bar{c}_{\beta},
	 d_{\al} \bar{d}_{\beta}
\right) $.

By substituting (\ref{eq:Zcyl}) into the equivalence of both
partition functions, eq.(\ref{def:equiv-dual}), and rewriting the
partition function of the open string picture with $S$-matrix, we obtain
\bea
 \label{eq:Z}
(M_{\al})_{\beta}^{\;\; j}
&=& 
(n_\al)_{\beta}^{\;\; i} S_i^{\;\; j}
.
\eea
Fusion rule coefficients and $S$ matrix have been given in
\cite{gab3,roh1,flohr1} and
$S$ undertakes two solutions, one of which is given from the other
 by flipping
the role of $R_{0,1}$. With the identification of $n_{\al}$ with
$\N_\al$, the fusion rule matrix, (\ref{eq:Z}) dramatically reduces the
potential 25 parameters of $M_\al$ to four:
\bea
 \label{sol:matrix}
  \ga &=& 1/\epsilon \;, \,\,\,\,
   a_{0,1} = \bar{a}_{0,1} = 0 ,
\nn
  c_0 &=& 2 c_{-\frac18} = 2 c_{\frac38} = \frac{4}{\bar{c}_0} = 
  \frac{2}{\bar{c}_{-\frac18}} = \frac{2}{\bar{c}_{\frac38}} \not= 0 ,
\nn
  d_0 &=& -2 d_{-\frac18} = -2 d_{\frac38} = -\frac{4}{\bar{d}_0}
  \frac{2}{\bar{d}_{-\frac18}} = \frac{2}{\bar{d}_{\frac38}} \not= 0,
\nn
  {a}_{-\frac18} &=& - {a}_{\frac38}
 =  \frac{i \epsilon}{2 \bar{a}_{-\frac18}}
 =  -\frac{i \epsilon}{2 \bar{a}_{\frac38}} \,\, ,
\eea
where the infinitesimal parameter $\epsilon$ is introduced to describe 
the solutions.
A general solution of boundary states can be easily obtained by
 substituting (\ref{sol:matrix}) into (\ref{def:boundary state})
 while the other solution of $S$ merely changes the sign of $\epsilon$
 in the last line.
The factor $\ga$ may be regarded as a regulator, taking a limit of
 $\epsilon\rightarrow 0$, and values of $a_\al, \bar a_\al$ are
  restricted by the limit, while other non-zero free parameters remain intact.
In other words, explicit form of solutions totally depends on how we
 take the limit in $a_{\al}$($\bar a_{\al}$) space.

It is easy to see that there are three kinds of limits, and,
 according to them, there arise three distinct
 families of solutions. In all cases, in both bra and ket state spaces,
 boundary conditions $\wt R_0, \wt R_1$ can neither be
   distinguished nor be excluded by the other and, therefore, we label
 them by $\wt R_c$ in both state spaces. Setting $c_0=d_0=2$ and
 $a_\al=b_\al$ for simplicity, it follows that 
\bea
 \label{sol:bs}
&(i)&   \left\{ \begin{array}{ll} 
   \bra{\wt R_c}=2\bra{-\frac18}+2\bra{\frac38} ,
 & \ket{\wt R_c}=2\ket{-\frac18}-2\ket{\frac38} \\[5pt]
\bra{\wt{-\frac18}}=
       \frac{i}{2} \bra{R} + \bra{-\frac18}-\bra{\frac38} ,
 & \ket{\wt{-\frac18}}=  \ket{\wt{\frac38}}=
       \ket{-\frac18}+\ket{\frac38}  \\[5pt]
   \bra{\wt{\frac38}}=
      -\frac{i}{2} \bra{R} + \bra{-\frac18}-\bra{\frac38} 
 &    ~~~~~~~~
   \end{array} \right.
\nn[5pt] 
&(ii)&   \left\{ \begin{array}{ll} 
   \bra{\wt R_c}=2\bra{-\frac18}+2\bra{\frac38} ,
 & \ket{\wt R_c}=2\ket{-\frac18}-2\ket{\frac38} \\[5pt]
   \bra{\wt{-\frac18}}= \bra{\wt{\frac38}}=
                             \bra{-\frac18}-\bra{\frac38} ,
 & \ket{\wt{-\frac18}}= 
       \frac{i}{2} \ket{R} + \ket{-\frac18}+\ket{\frac38} \\[5pt]
 & \ket{\wt{\frac38}}= 
      -\frac{i}{2} \ket{R} + \ket{-\frac18}+\ket{\frac38}  \\[5pt]
   \end{array} \right.
\nn[5pt]
&(iii)&   \left\{ \begin{array}{ll} 
   \bra{\wt R_c}=2\bra{-\frac18}+2\bra{\frac38} ,
 & \ket{\wt R_c}=2\ket{-\frac18}-2\ket{\frac38} \\[5pt]
   \bra{\wt{-\frac18}}=  \bra{\wt{\frac38}}=
                             \bra{-\frac18}-\bra{\frac38} ,
 & \ket{\wt{-\frac18}}= \ket{\wt{\frac38}}= 
                             \ket{-\frac18}+\ket{\frac38}
   \end{array} \right.
\nn
\eea
where $\ket{R}\equiv\sqrt2 \left( \ket{\Omega} + \ket{\phi}\right)$ and 
$\wt R_c$ is called the combined logarithmic boundary condition. 
Note that, with respect to the coefficients in front of `$R$' states in
 case (i) and (ii),
 only the relative sign of them is important, they can take any value in
  $\C$, unless we introduce another criteria to constrain them.

Most notably, these solutions indicate that we have either three states
for one end of the closed string tube and two for the other(case (i) and
(ii)), or two for each end respectively(case (iii)).

\subsubsection{First approach revisited}

Despite these interesting results, they should be discarded since final
states defined in (\ref{def:boundary state}) cannot be constructed by
the same way as initial states. A bra state $\bra{C,0}$ is not an upper
JLWV, but it generates $\bra{D,0}$ as $\bra{C,0}L_0 = h \bra{C,0} +\bra{D,0}$.
Thus, by the use of $\bra{C^*,0}=\bra{D,0}$, final states of
this case must be replaced by 
\bea
 \label{def:final state}
     \bra{\wt \al} &\equiv&  \sqrt{2}\, {a}_{\al} \bra{\omega}
   + \sqrt{2}\, {b}_{\al} \bra{\phi^*}
   + {c}_{\al} \bra{-\frac{1}{8}} + {d}_{\al} \bra{\frac{3}{8}},
\eea
where $\bra{\phi^*}$ is not the dual Ishibashi state of $\ket{\phi}$ but
of the lower JLWV of this cell.
In the bulk, these Ishibashi states of Jordan cell structures
do not propagate from initial boundary to final one and 
thus disappear from the partition function in the closed string picture. 
It follows that 
$\left(M_{\al}\right)_\beta^{-1/8}=c_{\al}\bar{c}_{\beta}$,
$\left(M_{\al}\right)_\beta^{3/8}=d_{\al}\bar{d}_{\beta}$,
others vanishing.
This is valid in any case of $c=-2$ LCFT which has been proposed so far.

In our first approach, fusion rules and
 this matrix cause a contradiction in (\ref{eq:Z}). Thus, we
 conclude that the above $n_{\al \beta}^{\;\; i}$ is not 
 the fusion rules given in \cite{gab3}. 
Leaving $n_{\al \beta}^{\;\; i}$ to be unknown, 
a part of (\ref{eq:Z}) shows $n_{\al \beta}^{\;\; 0} = n_{\al \beta}^{\;\; 1}$ and
 that the non-diagonal part of the $S^2$-matrix doesn't change the partition
 function. Thus, the theory remains invariant under this transformation. 
This means $S^2$ becomes effectively an unit matrix.

Now, (\ref{eq:Z}) reduces to
\bea
 \label{eq:cd-n}
   c_\al \bar{c}_\beta = 2 n_{\al \beta}^{\;\; 0} + n_{\al \beta}^{-1/8}, \quad
   d_\al \bar{d}_\beta = - 2 n_{\al \beta}^{\;\; 0} + n_{\al \beta}^{-1/8}, \quad
   n_{\al \beta}^{\;\; 0} = n_{\al \beta}^{\;\; 1}, \,\,
   n_{\al \beta}^{-1/8} = n_{\al \beta}^{3/8}.
\eea
Provided that $n_{\al \beta}^{\;\; i}$ is a positive integer and 
$\bar{c}_\al = c_\al^*$, $\bar{d}_\al = d_\al^*$, then the form of
(\ref{eq:cd-n}) already prescribes the solutions in three ways.
First, 
it prescribes the phase of coefficients, that is, if $c_\al\not= 0$ for some
$\al$, $\bar{c}_\beta$ has the opposite phase for an arbitrary
$\beta$, otherwise it vanishes. So, we can eliminate the phases without
loss of generality and set them to be real\footnote{
By setting $\beta=\al$, this means that $c_\al$ is a square root of
integer, thus, either integer or irrational.
}. 
Secondly, 
if ${}^\exists\al$; $c_\al \in \Z$ then ${}^\forall\beta$; $c_\beta \in
\Z$, and equivalently, 
if $\gamma$ is irrational and ${}^\exists\al$; $c_\al \in \gamma \Z$
then ${}^\forall\beta$; $c_\beta \in \gamma^{-1} \Z$. In other words, 
every coefficient must be integer or irrational of the same sort.
Lastly, rewriting eq.(\ref{eq:cd-n}), we draw another attention on the
coefficients as 
\bea
 \label{eq:cd-n-}
   c_\al c_\beta + d_\al d_\beta= 2 n_{\al \beta}^{-1/8} \geq 0, \quad
   c_\al c_\beta - d_\al d_\beta = 4 n_{\al \beta}^{\;\; 0} \geq 0, \quad
   n_{\al \al}^{-1/8} \geq 2 n_{\al \al}^{\;\; 0},
\eea
where the first prescription is used.
Note that there is no restrictions on $a_\al, b_\al$ and every possible
 boundary state has two additional degrees of freedom,
 since they give no contribution to naive inner products of
 boundary states. This point will be discussed later.

Collecting the above prescriptions, 
it becomes a simple task to pick up explicit solutions and several
of the simplest ones
are shown as below with a condition $n_{\al
\beta}^{\;\; i}\leq 4$. One may treat them as representatives.
\bea
 \label{sol:cd-n}
&(i)&   \left\{ \begin{array}{l} 
  \ket{\wt 0}= a_0 \ket{\Omega} + b_0 \ket{\phi} \\[5pt]
  \ket{\wt 1}= a_1 \ket{\Omega} + b_1 \ket{\phi} \\[5pt]
  \ket{\wt 2}= 2 \ket{-\frac18} \\[5pt]
  \ket{\wt 3}= 2 \ket{-\frac18} \pm 2 \ket{\frac38}
   \end{array} \right. , 
   \left\{ \begin{array}{l} 
  \left( n_{22}^{\;\; 0}, n_{22}^{-1/8} \right) = (1,2)  \\[5pt]
  \left( n_{33}^{\;\; 0}, n_{33}^{-1/8} \right) = (0,4)  \\[5pt]
  \left( n_{23}^{\;\; 0}, n_{23}^{-1/8} \right) = (1,2)  \\[5pt]
  {\rm other~} n_{\al \beta}^{\;\; 0} {\rm ~and~} n_{\al \beta}^{-1/8} {\rm ~vanish}
  \end{array} \right. , 
\nn[5pt] 
&(ii)&   \left\{ \begin{array}{l} 
  \ket{\wt 0}= a_0 \ket{\Omega} + b_0 \ket{\phi} \\[5pt]
  \ket{\wt 1}= a_1 \ket{\Omega} + b_1 \ket{\phi} \\[5pt]
  \ket{\wt 2}= 2 \ket{-\frac18} + 2 \ket{\frac38} \\[5pt]
  \ket{\wt 3}= 2 \ket{-\frac18} - 2 \ket{\frac38}
   \end{array} \right. , 
   \left\{ \begin{array}{l} 
  \left( n_{22}^{\;\; 0}, n_{22}^{-1/8} \right) = (0,4)  \\[5pt]
  \left( n_{33}^{\;\; 0}, n_{33}^{-1/8} \right) = (0,4)  \\[5pt]
  \left( n_{23}^{\;\; 0}, n_{23}^{-1/8} \right) = (2,0)  \\[5pt]
  {\rm other~} n_{\al \beta}^{\;\; 0} {\rm ~and~} n_{\al \beta}^{-1/8} {\rm ~vanish}
  \end{array} \right. , 
\nn[5pt] 
&(i_a)&    \left\{ \begin{array}{l} 
  \ket{\wt 0}= a_0 \ket{\Omega} + b_0 \ket{\phi} \\[5pt]
  \ket{\wt 1}= a_1 \ket{\Omega} + b_1 \ket{\phi} \\[5pt]
  \ket{\wt 2}= 2 \sqrt2 \ket{-\frac18} \\[5pt]
  \ket{\wt 3}= \sqrt2 \ket{-\frac18} \pm \sqrt2 \ket{\frac38}
   \end{array} \right. , 
   \left\{ \begin{array}{l} 
  \left( n_{22}^{\;\; 0}, n_{22}^{-1/8} \right) = (2,4)  \\[5pt]
  \left( n_{33}^{\;\; 0}, n_{33}^{-1/8} \right) = (0,2)  \\[5pt]
  \left( n_{23}^{\;\; 0}, n_{23}^{-1/8} \right) = (1,2)  \\[5pt]
  {\rm other~} n_{\al \beta}^{\;\; 0} {\rm ~and~} n_{\al \beta}^{-1/8} {\rm ~vanish}
  \end{array} \right. , 
\nn[5pt] 
&(ii_a)&    \left\{ \begin{array}{l} 
  \ket{\wt 0}= a_0 \ket{\Omega} + b_0 \ket{\phi} \\[5pt]
  \ket{\wt 1}= a_1 \ket{\Omega} + b_1 \ket{\phi} \\[5pt]
  \ket{\wt 2}= \sqrt2 \ket{-\frac18} + \sqrt2 \ket{\frac38} \\[5pt]
  \ket{\wt 3}= \sqrt2 \ket{-\frac18} - \sqrt2 \ket{\frac38}
   \end{array} \right. , 
   \left\{ \begin{array}{l} 
  \left( n_{22}^{\;\; 0}, n_{22}^{-1/8} \right) = (0,2)  \\[5pt]
  \left( n_{33}^{\;\; 0}, n_{33}^{-1/8} \right) = (0,2)  \\[5pt]
  \left( n_{23}^{\;\; 0}, n_{23}^{-1/8} \right) = (1,0)  \\[5pt]
  {\rm other~} n_{\al \beta}^{\;\; 0} {\rm ~and~} n_{\al \beta}^{-1/8} {\rm ~vanish}
  \end{array} \right. , 
\eea
where two Ishibashi states of Jordan cells in $\ket{\wt 2}$ and $\ket{\wt 3}$
are suppressed for simplicity and final boundary states are given by
the complex conjugation $*$, not by the dual $\dag$. 
Only representatives of nontrivial $n_{\al \beta}^{\;\; i}$ are listed. 
The solutions,
$(i_a)$ and $(ii_a)$ can be obtained from
the solutions, $(i)$ and $(ii)$, and vice versa, 
via redefinitions of boundary states,
which would be determined by the detailed analysis of some experiment, 
if it exists. 
Note that it is not necessarily two
in the solutions that have $\ket{-\frac18}$ and 
$\ket{\frac38}$, since there is at least one state of a Jordan cell 
with which we
can make another independent state. This will be discussed later.

It is possible to reduce the number of solutions by imposing another
condition, which is orthogonality of boundary states. For this
purpose, we must introduce the inner product which is
defined by Cardy in his paper \cite{card1},
\bea
 \label{def:innerprod-bs}
   \left( \al | \beta \right) \equiv \lim_{q\rightarrow 1} 
 \frac{\braket{\al|q^{L_0}|\beta}}{\left(\braket{\al|\al}
 \braket{\beta|\beta}\right)^{\frac12}} , 
\eea
as a bilinear form on boundary states. Although the states of
Jordan cells become undefined with themselves under this product, the
orthogonality of
states given below is still valid with arbitrary additions of those Ishibashi states.
Namely, it turns out that, among the solutions in 
eq.(\ref{sol:cd-n}), only $(ii)$ and $(ii_a)$ satisfy the
orthogonality as 
\bea
 \label{eq:ortho-bs}
    \left( \wt 2 | \wt 3 \right) = \lim_{q\rightarrow 1} 
 \frac{\braket{\wt 2|q^{L_0}|\wt 3}}{\left(\braket{\wt 2|\wt 2}
 \braket{\wt 3|\wt 3}\right)^{\frac12}}
 = \lim_{q\rightarrow 1} \frac{\braket{-\frac18|q^{L_0}|-\frac18} -
 \braket{\frac38|q^{L_0}|\frac38}}{\braket{-\frac18 | -\frac18} +
 \braket{\frac38 | \frac38}}
 = 0 .
\eea
The other orthogonality conditions hold trivially.
Hence, we may conclude that the simplest and most acceptable
solution is $(ii_a)$.
It is remarkable that this is the only solution which satisfies 
$n_{\al \beta}^{\;\; i}< 4$.

Simply following these prescriptions, the other approaches can be derived similarly.

\subsubsection{Second approach: on Flohr's construction}

In \cite{flohr1}, it was shown that there is a set of functions which
can provide
a modular invariant partition function and an explicit form of the $S$-matrix.
Subsequently, he generalised the set of functions and proposed three
different sets of functions and corresponding $S$-matrices, as 
case I, II and III. The case I is determined by requiring integer valued
fusion rules, while the case II and case III are by requiring symmetric
$S$-matrices and matching the characters calculated by the spectrum,
that is, keeping the original four $W$-characters unchanged.

Let us examine the case I. $S$-matrix of this case satisfies $S^2=1$ and
is parameterised by two complex numbers, $x$ and $y$,
which are related to each other. By setting $x=0$, $y=-1$, and relabeling the
characters as $\{\wt\chi_{1,2},\chi_{1,2},\chi_{-1,2},\chi_{0,2},\chi_{2,2}\}=
\{\chi_{-1},\chi_0,\chi_1,\chi_2,\chi_3\}$, we obtain similar equations
to (\ref{eq:cd-n}) from (\ref{eq:Z}) 
with $M_\al=(0,0,0,c_\al\bar{c}_\beta,d_\al\bar{d}_\beta )$, namely,
\bea
 \label{eq:cd-n1}
   c_\al \bar{c}_\beta
   = n_{\al \beta}^{\;\; 0} + n_{\al \beta}^{\;\; 2} , \quad
   d_\al \bar{d}_\beta
   = - n_{\al \beta}^{\;\; 0} + n_{\al \beta}^{\;\; 2} , \quad
   n_{\al \beta}^{-1} = 0, \,\,
   n_{\al \beta}^{\;\; 0} = n_{\al \beta}^{\;\; 1}, \,\,
   n_{\al \beta}^{\;\; 2} = n_{\al \beta}^{\;\; 3},
\eea
and as the third prescription,
\bea
 \label{eq:cd-n1-}
   c_\al c_\beta + d_\al d_\beta= 2 n_{\al \beta}^{\;\; 2} \geq 0, \quad
   c_\al c_\beta - d_\al d_\beta = 2 n_{\al \beta}^{\;\; 0} \geq 0, \quad
   n_{\al \al}^{\;\; 2} \geq n_{\al \al}^{\;\; 0}.
\eea
Because fusion rules in \cite{flohr1} have negative integer values,
it seems necessary to remove the assumption of positive $n_{\al \beta}^{\;\; i}$.
However, those fusion rules contradict this (\ref{eq:cd-n1}) and, thus,
there is no need to expect the identification $n_\al=\N_\al$. We may
keep the assumption.

It is easy to follow the same procedure and, if we require $n_{\al
\beta}^{\;\; i}<4$, we find three solutions, one of which is given by
multiplying the coefficients and $n_{\al \al}^i$ by a factor of $\sqrt2$ and
2, respectively. Two of them are
\bea
 \label{sol:cd-n1}
&(i)&   \left\{ \begin{array}{l} 
  \ket{\wt 0}= a_0 \ket{\Omega} + b_0 \ket{\phi} \\[5pt]
  \ket{\wt 1}= a_1 \ket{\Omega} + b_1 \ket{\phi} \\[5pt]
  \ket{\wt 2}= 2 \ket{-\frac18} \\[5pt]
  \ket{\wt 3}= \ket{-\frac18} \pm \ket{\frac38}
   \end{array} \right. , 
   \left\{ \begin{array}{l} 
  \left( n_{22}^{\;\; 0}, n_{22}^{\;\; 2} \right) = (2,2)  \\[5pt]
  \left( n_{33}^{\;\; 0}, n_{33}^{\;\; 2} \right) = (0,1)  \\[5pt]
  \left( n_{23}^{\;\; 0}, n_{23}^{\;\; 2} \right) = (1,1)  \\[5pt]
  {\rm other~} n_{\al \beta}^{\;\; 0} {\rm ~and~} n_{\al \beta}^{\;\; 2} {\rm ~vanish}
  \end{array} \right. , 
\nn[5pt] 
&(ii)&   \left\{ \begin{array}{l} 
  \ket{\wt 0}= a_0 \ket{\Omega} + b_0 \ket{\phi} \\[5pt]
  \ket{\wt 1}= a_1 \ket{\Omega} + b_1 \ket{\phi} \\[5pt]
  \ket{\wt 2}= \ket{-\frac18} + \ket{\frac38} \\[5pt]
  \ket{\wt 3}= \ket{-\frac18} - \ket{\frac38}
   \end{array} \right. , 
   \left\{ \begin{array}{l} 
  \left( n_{22}^{\;\; 0}, n_{22}^{\;\; 2} \right) = (0,1)  \\[5pt]
  \left( n_{33}^{\;\; 0}, n_{33}^{\;\; 2} \right) = (0,1)  \\[5pt]
  \left( n_{23}^{\;\; 0}, n_{23}^{\;\; 2} \right) = (1,0)  \\[5pt]
  {\rm other~} n_{\al \beta}^{\;\; 0} {\rm ~and~} n_{\al \beta}^{\;\; 2} {\rm ~vanish}
  \end{array} \right. ,
\eea
where only the second solution satisfies the orthogonality of the states.
The above results are valid for any $x$ and $y$ 
unless $(x+y)^2 + 2 (xy -1)^2 = 0$.
A point which should be mentioned is that these characters are not based
on analysis of primary fields and states, so it is not clear whether
there is a Jordan cell other than at $h=0$. 
In the first approach, an introduced
generalised hwrep contains a Jordan cell structure at $h=1$ and, even if
we assume them in this approach, nothing seems to be changed. So,
we add two linear combinations of states of Jordan cells 
as above, in order to show those possibilities. 

\subsubsection{Third approach: on Kogan and Wheater's construction}

This approach is on the basis of the first results which describe LCFT in
the presence of a boundary\cite{kog3}.
Various 2-point functions have been calculated and shown 
logarithmic singularities in boundary LCFT.
In the $c=-2$ case, since the discussion only deals with one Jordan cell
which appears in the calculated 2-point function, it may be reasonable to 
assume
that there is only one Jordan cell in the theory. We will take this
assumption for this approach.\footnote{
This does not totally exclude the possibility of a Jordan cell at $h=1$ and
it should be confirmed by the direct calculation of the 4-point
function, $\braket{\phi_{1,2}\phi_{2,2}\phi_{1,2}\phi_{2,2}}$, 
without boundaries.
}
Since our conjecture is that only one state survive in the cell, their
 construction, using both fields in the cell, contradicts
  ours. However, it would be interesting to take their set of characters
   in the open string picture and examine how they change the equations.
In addition, it is known that there is a representation of conformal
 dimension one, and that its character is $\chi_{\V_1}$, we then have to
 admit the states, $\ket{\phi}$ and $\bra{\phi^*}=\bra{\phi}$.
But, our prescriptions are still applicable, because they never
 appear in the boundary states due to the fact that
 $\chi_{\V_1}$ generates $\Delta\Theta_{1,2}/\eta$ under a modular
  transformation which is absent in this open string picture. 

The eq.(\ref{eq:Z}) and $M_\al$ lead to 
\bea
 \label{eq:cd-n4}
   c_\al \bar{c}_\beta
   = \frac12 n_{\al \beta}^{\;\; 0} + n_{\al \beta}^{\;\; 2} , \quad
   d_\al \bar{d}_\beta
   = - \frac12 n_{\al \beta}^{\;\; 0} + n_{\al \beta}^{\;\; 2} , \quad
   n_{\al \beta}^{\;\; 0} = n_{\al \beta}^{\;\; 1}, \,\,
   n_{\al \beta}^{\;\; 2} = n_{\al \beta}^{\;\; 3},
\eea
and the prescription is
\bea
 \label{eq:cd-n4-}
   c_\al c_\beta + d_\al d_\beta= 2 n_{\al \beta}^{\;\; 2} \geq 0, \quad
   c_\al c_\beta - d_\al d_\beta = n_{\al \beta}^{\;\; 0} \geq 0, \quad
   2 n_{\al \al}^{\;\; 2} \geq n_{\al \al}^{\;\; 0}.
\eea
Note that the above eq.(\ref{eq:cd-n4}) coincides with eq.(44) in
\cite{kog3}.
With a trick of defining $c_\al^\prime=\sqrt2 c_\al$,
$d_\al^\prime=\sqrt2 d_\al$,
solutions which satisfy $n_{\al \beta}^i < 4$ are easily found as
\bea
 \label{sol:cd-n4}
&(i)&   \left\{ \begin{array}{l} 
  \ket{\wt 1}= a_1 \ket{\Omega} \\[5pt]
  \ket{\wt 2}= \sqrt2 \ket{-\frac18} \\[5pt]
  \ket{\wt 3}= \sqrt2 \ket{-\frac18} \pm \sqrt2 \ket{\frac38}
   \end{array} \right. , 
   \left\{ \begin{array}{l} 
  \left( n_{22}^{\;\; 0}, n_{22}^{\;\; 2} \right) = (2,1)  \\[5pt]
  \left( n_{33}^{\;\; 0}, n_{33}^{\;\; 2} \right) = (0,2)  \\[5pt]
  \left( n_{23}^{\;\; 0}, n_{23}^{\;\; 2} \right) = (2,1)  \\[5pt]
  {\rm other~} n_{\al \beta}^{\;\; 0} {\rm ~and~} n_{\al \beta}^{\;\; 2} {\rm ~vanish}
  \end{array} \right. , 
\nn[5pt] 
&(ii)&   \left\{ \begin{array}{l} 
  \ket{\wt 1}= a_1 \ket{\Omega} \\[5pt]
  \ket{\wt 2}= \ket{-\frac18} + \ket{\frac38} \\[5pt]
  \ket{\wt 3}= \ket{-\frac18} - \ket{\frac38}
   \end{array} \right. , 
   \left\{ \begin{array}{l} 
  \left( n_{22}^{\;\; 0}, n_{22}^{\;\; 2} \right) = (0,1)  \\[5pt]
  \left( n_{33}^{\;\; 0}, n_{33}^{\;\; 2} \right) = (0,1)  \\[5pt]
  \left( n_{23}^{\;\; 0}, n_{23}^{\;\; 2} \right) = (2,0)  \\[5pt]
  {\rm other~} n_{\al \beta}^{\;\; 0} {\rm ~and~} n_{\al \beta}^{\;\; 2} {\rm ~vanish}
  \end{array} \right. ,
\eea
where, again, only the second solution satisfies the orthogonality. 
Notably, these solutions are different from what is shown in
\cite{kog3}, due to our assumption and prescriptions.
Precisely speaking, the second prescription directly prohibits us from having
 those solutions with different types of irrational coefficients. 
All the prescriptions are derived from the
assumption that $n_{\al \beta}^{\;\; i}$ is positive integer and
$\bar{c}_\al=c_\al$, $\bar{d}_\al=d_\al$, so that
the second prescription eliminates such solutions. Thus, we may recover
 their solution by changing the assumption of $n_{\al \beta}^{\;\; i}$.
Their solution was derived with a condition, $n_{\al \al}^{\;\; 2}\leq 1$, 
in order to look for the first simplest example. 
It was also one of the differences from ours.

\subsubsection{Conclusions \& Remarks on approaches}

To summarise, we make some general remarks on all the solutions shown in
this section,
setting $\chi_2\equiv\chi_{-\frac18}$
and $\chi_3\equiv\chi_{\frac38}$ in the first approach.

First, all the solutions have 
\bea
 \label{sol:com999}
n_{\al \beta}^{\;\; 0}=n_{\al \beta}^{\;\; 1}
 ~~and~~ n_{\al \beta}^{\;\; 2}=n_{\al \beta}^{\;\; 3}
\eea
in common as a part of solutions.
Secondly, it must be mentioned that 
it is possible to have three boundary states whose coefficients of
$\ket{-\frac18}$ or $\ket{\frac38}$ take non-zero values simultaneously.
For instance, in solutions $(i)$, if we split 
$\ket{\wt 3}$ to $\ket{\wt 3_\pm}$ with positive and negative signs in its
expressions, we can have such states as $\ket{\wt 2}$, $\ket{\wt 3_+}$ and
$\ket{\wt 3_-}$.
Thirdly, 
amongst all the solutions, we conclude that the most acceptable
solutions are illustrated with 
$(ii)$ (or $(ii_a)$) because of the orthogonality. Most strikingly,
all approaches have the following orthogonal solution,
\bea
 \label{sol:cd-n999}
    \left\{ \begin{array}{l} 
  \ket{\wt 0}= a_0 \ket{\Omega} + b_0 \ket{\phi} \\[5pt]
  \ket{\wt 1}= a_1 \ket{\Omega} + b_1 \ket{\phi} \\[5pt]
  \ket{\wt 2}= \sqrt2 \ket{-\frac18} + \sqrt2 \ket{\frac38} \\[5pt]
  \ket{\wt 3}= \sqrt2 \ket{-\frac18} - \sqrt2 \ket{\frac38}
   \end{array} \right. , 
   \left\{ \begin{array}{l} 
  \left( n_{22}^{\;\; 0}, n_{22}^{\;\; 2} \right) = (0,2)  \\[5pt]
  \left( n_{33}^{\;\; 0}, n_{33}^{\;\; 2} \right) = (0,2)  \\[5pt]
n_{23}^{\;\; 2} = 0
  \end{array} \right. , 
\eea
where, of course, whether there are both $\ket{\wt 0}$ and $\ket{\wt 1}$
depends on how many Jordan cells exist in the theory. In this solution,
only $n_{23}^{\;\; 0}=n_{23}^{\;\; 1}=n \in\Z^+$ has a dependency on approaches and it
is caused by the difference of definitions of characters, $\chi_0$
and $\chi_1$.

In the first and second approaches, although fusion rules are defined in
the original constructions, they cannot be identified with 
$n_{\al \beta}^{\;\; i}$ and an interpretation of $n_{\al \beta}^{\;\; i}$ as
fusion rules is excluded at this point.
Besides, when we look at the solution (\ref{sol:com999}), it is obvious that we
miss an appropriate $n_{\al \beta}^{\;\; i}$ s.t. 
$n_{\al \beta}^{\;\; i}=\delta_{\beta}^{i}$
and lose such a way to the Verlinde formula as in boundary 
unitary CFTs. 

In the closed string picture, whatever boundary states we have, those of
Jordan cell cannot travel from one end to the other while those which 
have nontrivial coefficients of $\ket{-\frac18}$ and $\ket{\frac38}$ can
contribute to the partition function. With the definition of inner
product of boundary states, Ishibashi states of the cell become null, 
hence one
might think that those states of the cell can be regarded as `ghosts'
in this picture.

In the open string picture, 
in spite of the vanishing physical degrees of freedom of the cell in the
other picture, there is non-zero $n_{\al \beta}^{\;\; 0}=n_{\al
\beta}^{\;\; 1}\in \Z^+$. It means that those indecomposable representations can
travel along the edges of the strip and appear in the partition
function.
It would be interesting to check this result in the context of
condensed matter physics.

It should be noted that, if the theory on the annulus is the
non-chiral theory in \cite{gab5}, our results become inapplicable,
because logarithmic fields of the theory are expected to satisfy 
the condition (\ref{def:bc0}) by definition.
However, this doesn't mean our construction is totally invalid for
all non-chiral theories, though (\ref{calc}) should be
modified in some proper way.
For instance, by imposing an appropriate restriction on bra states in (\ref{calc}).

\section{Summary and Discussions}

On the basis of the mathematical definition of JLWM, we have proven that 
there exists the Ishibashi state in a $rank$-2 Jordan cell structure
and only one is allowed in the structure.
We have also shown the explicit form of it in terms of primary states
in the Hilbert space of the theory. 
The result is that descendants of the normal primary state can be in the
expression of Ishibashi state, but those of the logarithmic state are
excluded.
We conjecture that this holds for all LCFTs which 
contain, at least, one $rank$-2 JLWM as a submodule.
It is prominently useful in a sense that, while Ishibashi states of diagonal
representations are given conventionally, 
the whole set of such states, including those of Jordan cells,
provide a general expression of boundary states. 
Therefore, it leads to relations between boundary
states and $n_{\al \beta}^i$ and, hopefully, the Verlinde formula.

In the previous section, we have focused on the $c=-2$ theory as an example.
Introducing the $c=-2$ theory, its characters and $S$-matrices, 
we have constructed partition functions of the boundary LCFT, and derived
boundary states and the relations of $n_{\al \beta}^i$.
On the way to construct the partition function, there are two choices
of final boundary states and we have discussed both and excluded the
former choice because of the condition (\ref{def:bc0}) and the
definition of Jordan 
cell structure. Both choices finally turned out not to formulate any
conventional expression of the Verlinde formula.

Still, it is interesting to interpret the results in the D-brane context,
since boundary states are the initial and final states of closed
strings, and some of them would correspond to those on the D-branes.
Here, we should recall that we do not know 
the clear microscopic interpretation of them
in such a trivial manner as 
 the free and (anti-)symmetric boundary conditions in Ising model. 
So, the interpretation is still obscure.

As for the Verlinde formula,
in spite of the construction in \cite{flohr1} that some of $S$-matrices 
satisfy it and give integer values of fusion coefficients,
our results indicate that they never give the
conventional formulas as in boundary unitary CFT.
The results also imply that it is impossible to have such a
formula or, at least, necessary to modify its expression. 
On the other hand, with characters in \cite{gab3,roh1}, 
it was shown 
that they do not satisfy the conventional one but lead to a block-diagonal 
form, which no longer expresses fusion
matrices only with S-matrix. Even this case is not applicable to
our results
because our $n_{\al \beta}^{\;\; i}$
do not match the fusion rules. 
After all, we lost the complete identification of
$n_{\al}$ and $\N_\al$, and it suggests that another criteria should be
introduced to describe the theory in the open string picture. 
Apart from it, questions still remain, what else can
$n_{\al}$ be and whether fusion rules of the open
string picture is different from those of ordinary CFT.
It would be interesting to answer these questions.

Recently, in \cite{kaw}, another attempt has been done
in order to construct boundary states of $c=-2$
rational LCFT from the $(\xi, \eta)$-ghost system given in
\cite{kau2,salu}. By setting $n_{\wt 0 \,i}^{\;\;j}=\delta_i^j$, 
they suspect that it is possible to make Ishibashi states from the ghost
system, and show that it is impossible to derive
the Verlinde formula in a conventional way.
This also supports our observation of incompatibility between Ishibashi
states and the fusion rules. So, it is also interesting to confirm
whether we can construct our Ishibashi states from the $(\xi,
\eta)$-ghost or symplectic fermion system\cite{kau2,kau3}, or from any
other field representation\cite{gura,kog2} with our $n_{\al \beta}^{\;\;i}$.

After this paper was completed, one refference was added to the end of
\cite{rahimi1}, which also deals with boundary LCFT.

\vspace{30pt}\noindent
{\Large {\bf Acknowledgments}}\\[-5pt]

The author would like to thank I.I. Kogan
for suggesting this problem, stimulating discussions and careful
reading of the manuscript,
and J.F. Wheater and S. Kawai for useful discussions.
Y.I.
would also like to thank A.Nichols for his interesting discussions
and suggestions to the manuscript.

\end{document}